\documentclass{elsart}
\makeatletter
\let\AND\original@AND
\makeatother
\usepackage{amssymb}
\usepackage{amsmath}
\usepackage{color}
\usepackage{curves}
\usepackage{eepic}
\usepackage{epic}
\usepackage{graphics,color}
\usepackage{pstricks}
\usepackage{subfigure}
\usepackage{graphics}
\usepackage{pstricks-add,pst-node}
\usepackage{algorithm}
\usepackage{algorithmic}
\usepackage{pifont}
\newtheorem{pr}{Property}

\newtheorem{con}{Conjecture}
\newtheorem{qu}{Question}

\newcommand{\Nb}{\mathbb{N}}
\newcommand{\val}{\text{val}}
\usepackage{enumitem}

\usepackage{tikz}
\usetikzlibrary{arrows,automata,positioning}
\usetikzlibrary{calc}
\tikzstyle{vertex} = [circle, draw, thick, text centered]

\begin{document}

\begin{frontmatter}

  \title{Optimal Adjacent Vertex-Distinguishing Edge-Colorings of Circulant Graphs}

\date{}

 \author[UJF]{Sylvain Gravier}   \author [UJF,ENS] {Hippolyte Signargout} \author[USTHB]{Souad Slimani}
   \address[UJF]{UGA-CNRS, Institut Fourier, FED Maths \`a Modeler, 100 rue des maths, BP 74, 38402 St Martin d'H\`eres cedex, France}
   \address[USTHB]{Laboratoire LaROMaD, FDR Maths \`a Modeler. U.S.T.H.B Universit\'e, Facult\'e des Math\'ematiques
BP 32 El Alia Bab Ezzouar 16 111 Alger}
\address[ENS]{ENS de Lyon, Département d'Informatique}

\begin{abstract}
A $k$-proper edge-coloring of a graph $G$ is called \textit{adjacent vertex-distinguishing} if any two adjacent vertices are distinguished by the set of colors appearing in the edges incident to each vertex. The smallest value $k$ for which $G$ admits such coloring is denoted by $\chi^\prime _a(G)$. We prove that $\chi^\prime _a(G)=2R+1$ for most circulant graphs $C_n([\![1,R]\!])$. 

\end{abstract}

\end{frontmatter}
\section{Introduction}\label{sec1}
\noindent Let $G=(V,E)$ be a simple, finite, connected and undirected graph and $C$ be a set of colors. We use $N_G(v)$ (or simply $N(v)$ if the context makes it clear) to denote the set of neighbors of the vertex $v$ in $G$ and $d_G(v)$ (or $d(v)$) to denote the degree of $v$ in $G$. We call $\Delta(G)$ the maximum degree of $G$. An \textit{edge-coloring} $\varphi : E \rightarrow C$ is an assignment of colors to the edges of $G$. The mapping $\varphi$ is a $k$-\textit{edge-coloring} if $|C|=k$. It is said to be \textit{proper} if any two edges incident to the same vertex are mapped to different colors. The chromatic index, denoted $\chi^\prime(G)$, of a graph $G$ is the smallest integer $k$ such that $G$ admits a $k$-edge-coloring.

A proper edge-coloring $\varphi$ is \textit{adjacent vertex-distinguishing} (\textit{AVD-coloring} for short) if for any $(u,v)\in E$, $\varphi(N(u))\neq \varphi(N(v))$. The smallest number of colors required for an AVD-coloring of $G$ is called the AVD-chromatic index of $G$ and denoted $\chi_a^\prime(G)$.

Some interest has been shown in non-proper adjacent vertex-distinguishing edge-colorings. In \cite{AignerM1992Iaav} it was shown that the minimum number of colors necessary for a non-proper vertex-distinguishing edge-colorings is equivalent to $n^\frac{1}{k}$ for $k$-regular graphs of order $n$. However, a much stronger interest has been shown in proper adjacent vertex-distinguishing edge-colorings, which are thus most often only referred to as \textit{adjacent vertex-distinguishing}.

One may also read the term \textit{adjacent strong edge-coloring} or \textit{1-strong edge-colorings} as an effort to avoid confusion. The latter is part of the wider concept of $d$-\textit{strong edge-coloring}, which is a proper coloring in which two vertices at distance lower than $d$ of each other can not share the same set of incident colors. This concept introduced in \cite{AkbariS2006reco} also embraces (when $d$ is larger than the diameter of the graph) \textit{strong edge-colorings} (or \textit{vertex-distinguishing edge-colorings}), in which any two vertices do not share the same set of incident colors, whatever their distance.

Strong edge colorings were introduced independently by Fouquet et al. \cite{Fouquet1,Fouquet2} for radio networks and the frequency assignment problem and for graphs in \cite{FavaronOdile1996Seco}. They were deeply studied in \cite{BurrisA.C.1997VPE} where the conjecture was made that any graph of order $n\geq3$ with no isolated edge and at most one isolated vertex admitted an $(n+1)$-strong edge-coloring and in \cite{BazganCristina1999OtVP} where the conjecture was proven. The smallest number of colors used to obtain a such coloring, denoted by $\chi_S^\prime$, is called \textit{strong edge chromatic index} or \textit{observability}. It was studied for planar graphs in \cite{Chang}, and additional properties using odd graphs  were found in \cite{Wang} and extended to multigraphs in \cite{BalisterP.N.2003Veco}.

Trivial bounds on the AVD chromatic index of a graph $G$ are  $\Delta(G)\leq\chi^{\prime}(G)\leq  \chi^{\prime}_{a}(G)\leq\chi^{\prime}_{S}(G)$. In \cite{ZhangZf2002Asec}, the following conjecture was proposed :

\begin{con}\cite{ZhangZf2002Asec}
If $G$ is a simple connected graph on at least 3 vertices and $G\neq C_5$ (a cycle of order 5) then $\Delta(G)\leq \chi^\prime_{a}(G)\leq \Delta(G)+2$.
\end{con}

In \cite{BalisterP.N.2007AVDE}, it is shown that $\chi^\prime_{a}(G)\leq 5$ for all graph $G$ with maximum degree $\Delta(G)=3$ and $\chi^\prime_{a}(G)\leq \Delta(G)+2$ for bipartite graphs. For $k$-chromatic graphs $G$ without isolated edges, the authors proved that $\chi^\prime_{a}(G)=\Delta(G)+O(\log k)$. The study in \cite{BarilJean-Luc2006Avde} was done for Meshes graphs (the Cartesian product of $p$ paths and of $p$ cycles). In \cite{Baril_vertexdistinguishing}, it was shown that for any integers $n\geq 2$ and $d\geq 2$, we have $\chi^\prime_a(K_n^d)=\Delta(K_n^d)+1$, where $K_n^d$ is the Cartesian product of the complete graph $K_n$ by itself $d$ times (also known Hamming graph) and that for direct products of graphs, $\chi^\prime_{a}(G\times H)\leq min\{\chi^\prime(G).\chi_{a}(H), \chi^\prime_{a}(G).\chi^\prime(H)\}$. For graphs with maximum degree $\Delta\geq5$ and maximum average degree $\text{mad}<3-\frac2\Delta$, the result $\chi^{\prime}_a\leq\Delta+1$ is proved in \cite{HocquardHerve2013Avec} based on results of \cite{WangWeifan2010Avde}. A weaker bound of $\chi^{\prime}_a\leq\Delta+300$ was proved in \cite{HatamiHamed2005d+3i} for all graphs with $\Delta$ large enough.

For $n\in\mathbb{N}^*$ and $S\subset\mathbb{Z}_n$, the \textit{circulant graph} $C_n(S)$ is the non-directed graph whose $n$ vertices are the elements of $\mathbb{Z}_n$ with an edge $(i,j)$ if and only if $|i-j| \in S$. In this paper we will say two vertices $i$ and $j$ are at distance $d$ and that an edge $(i,j)$ is of length $d$ if $|i-j|=d$. We write $[\![a,b]\!]=\{i\in\mathbb{N}|a\leq i\leq b\}$.

The notion of $d$-strong edge-colorings of circulant graphs has been studied in \cite{Mockov}. An exact value of $\chi_d$ of $C_n(\{1,2\})$ is given for $d\in\{1,2\}$. We will use an idea similar to the one they introduced of cutting colored circulant graphs and merging them together to get colorings for graphs of higher order.

Obviously for a $k$-regular graph we have $\chi^{\prime}_a\geq k+1$. In this paper we will show that for most $R$ and most $n$, $\chi^{\prime}_a(C_n([\![1,R]\!]))=\Delta(C_n([\![1,R]\!]))+1$:

\begin{thm}\label{thm:all}
Let $R\geq 1$ and $n\geq(R+2)2^{1+\lfloor \log R \rfloor}\left(R+(R+2)2^{\lfloor \log R \rfloor}\right)-2R$. If $R\neq1 \mod6$ or $n=0\mod3$ then $\chi^{\prime}_a(C_n([\![1,R]\!]))=2R+1$.
\end{thm}

The main idea behind the proof of Theorem \ref{thm:all} is the following. We build optimal (in the number of colors used) adjacent vertex-distinguishing edge-colorings of $C_{mp}([\![1,R]\!])$ for some $m$ and all $p\in\Nb$ and of $C_{k}([\![1,R]\!])$ for some $k$ prime with $m$ such that both can be merged together to form a the graph $C_{mp+k}([\![1,R]\!])$ and a coloring for it. Then by varying $p$ and merging multiple times we can reach an optimal coloring of $C_{n}([\![1,R]\!])$ for any $n$ large enough. As in general $k<<mp$ we also call the merging operation \textit{adding an extension}. The optimal adjacent vertex-distinguishing edge-colorings of the extendable graph and of the extension are defined in Section \ref{sec:pre}. The process of adding extensions is described in Section \ref{sec:fun} to complete the proof of Theorem \ref{thm:all}. We conclude the paper in Section \ref{sec:per} with perspectives for future work.

\section{Preliminary Results}\label{sec:pre}

We first define an optimal proper edge-coloring of circulant graphs with only edges of odd length (Lemma \ref{lem:bas}). We also define a proper edge-coloring with only one more color and the additional property of distinguishing vertices close to each other (Lemma \ref{lem:str}). Both colorings are periodic, and can be applied only when the order of the graph is a multiple of their periods, but we show that it is possible to extend the coloring of Lemma \ref{lem:bas} to any even order by adding $2$-vertices extensions (Lemma \ref{lem:bet}).

We use these colorings to build a periodic and optimal AVD coloring of $C_{mp}([\![1,R]\!])$ for all $p\in\Nb$ where $m$ is the period of the coloring, which depends on $R$ (Lemma \ref{lem:las}). The edges of the graph can be partitioned so that the edges of each set induce a circulant graph with only edges of odd length. Each of these subgraphs can be colored with the colorings of Lemmas \ref{lem:bas}, \ref{lem:str} and \ref{lem:bet}. A careful choice of colors can ensure that the resulting coloring is proper, that vertices which are adjacent in $C_{mp}([\![1,R]\!])$ can be distinguished by the colors of the edges on which the coloring of Lemma \ref{lem:str} is applied, and that no such coloring can be obtained with less colors. 

We then define the extension by building vertex-distinguishing colorings for complete graphs of odd order (Property \ref{pr:cliq}), which can be seen as circulant graphs: $K_{2R+1}=C_{2R+1}([\![1,R]\!])$.

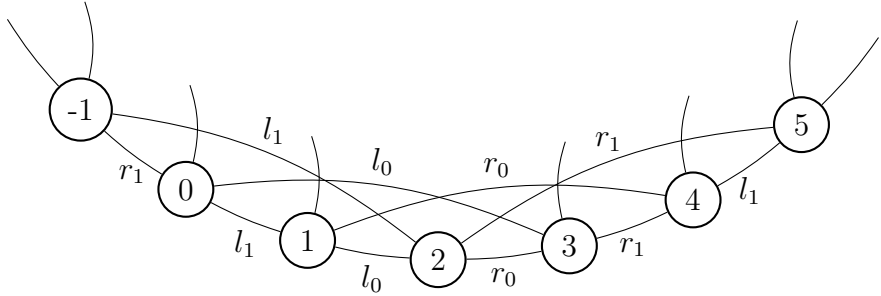
\begin{figure}[h]
    \centering
    \begin{tikzpicture}
    \foreach \x in {-1,...,5}
    \node[vertex] (\x) at (\x*360/25-120:7) {\x};
    \foreach \x in {-1,...,5}
    \node[vertex, draw=none] (t\x) [above=of \x] {};
    \node[vertex, draw = none] (-2) at (-2*360/25-120:7) {};
    \node[vertex, draw = none] (6) at (6*360/25-120:7) {};

    \path[-] 
    (-1)  edge [bend right=5, align=left, below] node  {$r_1$} (0);
    
    \path[-](0)  edge [bend right=5, align=left, below] node  {$l_1$} (1);
    \path[-](1)  edge [bend right=5, align=left, below] node  {$l_0$} (2);
    \path[-](2)  edge [bend right=5, align=left, below] node  {$r_0$} (3);
    \path[-](3)  edge [bend right=5, align=left, below] node  {$r_1$} (4);
    \path[-](4)  edge [bend right=5, align=left, below] node  {$l_1$} (5);
    \path[-](-2)  edge [bend right=5, align=left, below] node  {} (-1);
    \path[-](5)  edge [bend right=5, align=left, below] node  {} (6);
    \path[-](-1)  edge [bend left=15, align=left, above] node  {$l_1$} (2);
    \path[-](0)  edge [bend left=15, align=left, above] node  {$l_0$} (3);
    \path[-](1)  edge [bend left=15, align=left, above] node  {$r_0$} (4);
    \path[-](2)  edge [bend left=15, align=left, above] node  {$r_1$} (5);
    \foreach \x in {-1,0,1}
    \path[-](\x)  edge [bend right=15, align=left, above] node  {} (t\x);
    \foreach \x in {3,4,5}
    \path[-](\x)  edge [bend left=15, align=left, above] node  {} (t\x);
    
    \end{tikzpicture}
    \caption{$\varphi_2$ on a part of $C_{4k}(U_1)$, $k\geq2$}
    \label{fig:propcol}
\end{figure}

For $m\in \mathbb{N}$, let $U_m=\{2p+1|0\leq p\leq m\}$. Let $k \geq2$ and $v_0\in[\![0,2k(m+1)]\!]$. We name $\varphi_{v_0}$ the following edge-coloring of $C_{2(m+1)k}(U_m)$ using colors in $\{l_i|i\in[\![0,m]\!]\}\cup\{r_i|i\in[\![0,m]\!]\}$ (see Figure \ref{fig:propcol} for an illustration with $m=1$) :
\begin{itemize}
    \item For $i\in [\![0,m]\!]$, $\varphi_{v_0}(v_0+i,v_0+i+1)=r_i$ and $\varphi_{v_0}(v_0-i-1,v_0-i)=l_i$;
    \item For any vertex $v$, $\varphi_{v_0}(v+2m+2,v+2m+3)=\varphi_{v_0}(v,v+1)$;
    \item For any vertex $v$ and $p\leq m$, $\varphi_{v_0}(v-p,v+1+p)=\varphi_{v_0}(v,v+1)$.
\end{itemize}

We write $\varphi$ when the choice of $v_0$ is of no importance.

\begin{lem}\label{lem:bas}
For $m,k$ integers with $k\geq2$, $\varphi$ is a $2(m+1)$-proper coloring of $C_{2(m+1)k}(U_m)$.
\end{lem}

\begin{pf}

In this construction, any colors and any vertices can be swapped without changing the structure of the graph and coloring, so the proof that the coloring is proper can be reduced to proving that one vertex does not have the same color twice in its edge neighbourhood. For $m,k$ integers with $k\geq2$, as $C_{2(m+1)k}(U_m)$ is $(2m+2)$-regular, it is the same as showing that one vertex has $2m+2$ distinct colors in its edge neighbourhood, which can be seen by writing all colors in $S(v_0)$: for $p\in[\![0,m]\!]$, $\varphi(v_0,v_0+2p+1)=\varphi(v_0+p,v_0+1+p)=r_{p}$ and $\varphi(v_0-(2p+1),v_0)=\varphi(v_0-1-p,v_0-p)=l_{p}$. \qed
\end{pf}

By adding the color $0$ to the previous coloring and using a similar structure, we manage to distinguish the sets of colors of edges incident to nearby vertices. For a vertex $v_0$ of $C_{(2m+3)k}(U_m)$, we name $\phi_{v_0}$ the following edge-coloring:
\begin{itemize}
    \item $\phi_{v_0}(v_0,v_0+1)=0$;
    \item For $i\in [\![0,m]\!]$, $\phi_{v_0}(v_0+1+i,v_0+2+i)=r_i$ and $\phi_{v_0}(v_0-i-1,v_0-i)=l_i$;
    \item For any vertex $v$, $\phi_{v_0}(v+2m+3,v+2m+4)=\phi_{v_0}(v,v+1)$;
    \item For any vertex $v$ and $p\leq m$, $\phi_{v_0}(v-p,v+1+p)=\phi_{v_0}(v,v+1)$.
\end{itemize}

\begin{lem}\label{lem:str}
For $m,k$ integers with $k\geq2$, $\phi$ is a $(2m+3)$-proper edge-coloring of $C_{(2m+3)k}(U_m)$ which verifies:
\center
For two vertices $i$ and $j$, if $|i-j|\in[\![1,2m+2]\!]$ then $\phi(N(i))\neq\phi(N(j))$.
\end{lem}

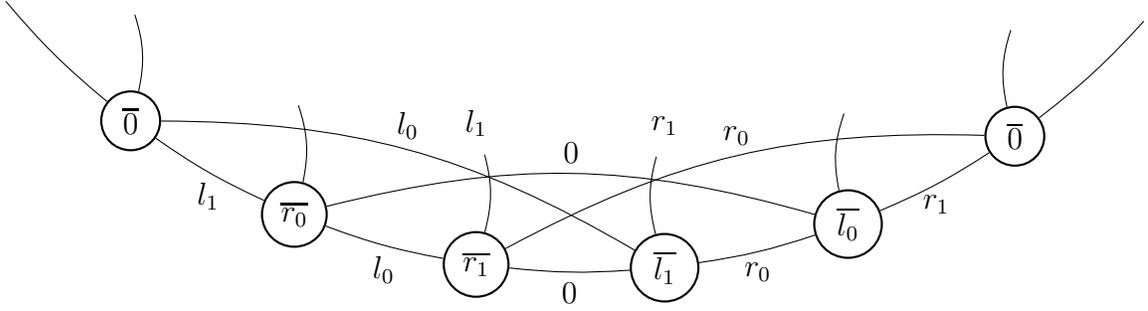
\begin{figure}[h]
    \centering
    \begin{tikzpicture}
    \node[vertex] (0) at (0*360/25-127:10) {$\overline 0$};
    \node[vertex] (1) at (1*360/25-127:10) {$\overline{r_0}$};
    \node[vertex] (2) at (2*360/25-127:10) {$\overline{r_1}$};
    \node[vertex] (3) at (3*360/25-127:10) {$\overline {l_1}$};
    \node[vertex] (4) at (4*360/25-127:10) {$\overline {l_0}$};
    \node[vertex] (5) at (5*360/25-127:10) {$\overline 0$};
    
    \foreach \x in {0,1,4,5}
    \node[vertex, draw=none] (t\x) [above=of \x] {};
     \node[vertex, draw=none] (t2) [above=of 2] {$l_1$};
      \node[vertex, draw=none] (t3) [above=of 3] {$r_1$};
    \node[vertex, draw = none] (-1) at (-1*360/25-127:10) {};
    \node[vertex, draw = none] (6) at (6*360/25-127:10) {};

    \path[-](0)  edge [bend right=5, align=left, below] node  {$l_1$} (1);
    \path[-](1)  edge [bend right=5, align=left, below] node  {$l_0$} (2);
    \path[-](2)  edge [bend right=5, align=left, below] node  {$0$} (3);
    \path[-](3)  edge [bend right=5, align=left, below] node  {$r_0$} (4);
    \path[-](4)  edge [bend right=5, align=left, below] node  {$r_1$} (5);
    
    \path[-](5)  edge [bend right=5, align=left, below] node  {} (6);
    \path[-](-1)  edge [bend right=5, align=left, above] node  {} (0);
    \path[-](0)  edge [bend left=15, align=left, above] node  {$l_0$} (3);
    \path[-](1)  edge [bend left=15, align=left, above] node  {$0$} (4);
    \path[-](2)  edge [bend left=15, align=left, above] node  {$r_0$} (5);
    \foreach \x in {0,1,2}
    \path[-](\x)  edge [bend right=15, align=left, above] node  {} (t\x);
    \foreach \x in {3,4,5}
    \path[-](\x)  edge [bend left=15, align=left, above] node  {} (t\x);
    
    \end{tikzpicture}
    \caption{$\phi$ on a part of $C_{(2m+3)k}(U_m)$ for $m=1$, $k\geq2$}
    \label{fig:firstr}
\end{figure}

\begin{pf}
Let $m,v_0,k$ be integers with $k\geq2$. In $C_{(2m+3)k}(U_m)$ the colors incident to $v_0-m-1$ are:
\begin{align*}\phi_{v_0}(v_0-m-1,v_0-m+2p)&=\phi_{v_0}(v_0+p-m-1,v_0+p-m)\\&=l_{m-p}\end{align*}
and \begin{align*}\phi_{v_0}(v_0-m-2-2p,v_0-m-1)&=\phi_{v_0}(v_0-m-2-p,v_0-m-1-p)\\&=\phi_{v_0}(v_0+1+m-p,v_0+2+m-p)\\&=r_{m-p}\end{align*}
for all $p\in[\![0,m]\!]$. The only color used by $\phi_{v_0}$ which is not in $S(v_0-m-1)$ is 0. We will say this vertex is a $\overline{0}$. The same way, for $i \in [\![0,m]\!]$, $v_0-i$ is a $\overline{r_{m-i}}$ and $v_0+i+1$ is a $\overline{l_{m-i}}$, as the vertices are labelled in Figure \ref{fig:firstr}. As the coloring is $(2m+3)$-periodic, this proves the two properties in the Lemma. First, the fact each vertex has $2m+2$ distinct colors in its edge neighbourhood shows the coloring is proper. Also, two vertices with the same set of colors in their edge neighbourhood are at a distance of at least $2m+3$, so vertices at distance lower than $2m+2$ are distinguished. \qed
\end{pf}

This property of $\phi$ covers the \textit{adjacent vertex-distinguishing} property while adding that vertices at even distance are also distinguished by the colors of their incident edges.

The proper coloring $\varphi$ of Lemma \ref{lem:bas} can only be applied when the order of the circulant graph $n$ is a multiple of $2(m+1)$. We can then build a coloring for a graph of order $n+2$ by adding a $2$-vertices extension to $\varphi$, and build a coloring for any even order by adding multiple extensions. The same strategy is used to prove Theorem \ref{thm:all} in Section \ref{sec:fun} with larger extensions. 

\begin{lem}\label{lem:bet}
For $m\in \mathbb{N}^*$ and $n\geq (2m+1)m$, $\chi^\prime\left(C_{2n}(U_m)\right)=2(m+1)$.
\end{lem}

\begin{pf}
Let $m\in \mathbb{N}^*$. We will first describe the extension on $C_{4m+4}(U_m)$ colored with $\varphi_{m+1}$ and then compute how many extensions are needed to color $C_{2n}(U_m)$ for $n\geq (2m+1)m$.
\begin{enumerate}
    
    \item \label{step} \begin{itemize}
        \item Add vertices $x$ and $y$;
        \item Add edges $(x,y)$, $(x,m+1)$ and $(y,m+2)$;
        \item Add edges $(m+1-2p,x)$, $(x,m+1+2p)$, $(m+2-2p,y)$ and $(y,m+2+2p)$ for $p\in[\![1,m]\!]$.
    \end{itemize} 
    \item \label{st:1} Remove the $2m+1$ edges of length $2m+1$ with one vertex on each side of the edge $(m+1,m+2)$. The colors of these edges are distinct: $(l_i)_{0\leq i\leq m-1}$ and $(r_i)_{0\leq i\leq m}$. The resulting graph is isomorphic to $C_{4m+6}(U_m)$, as can be seen by placing $x$ and $y$ between $m+1$ and $m+2$ in the way shown in Figure \ref{fig:fstext} for $m=2$.
    
    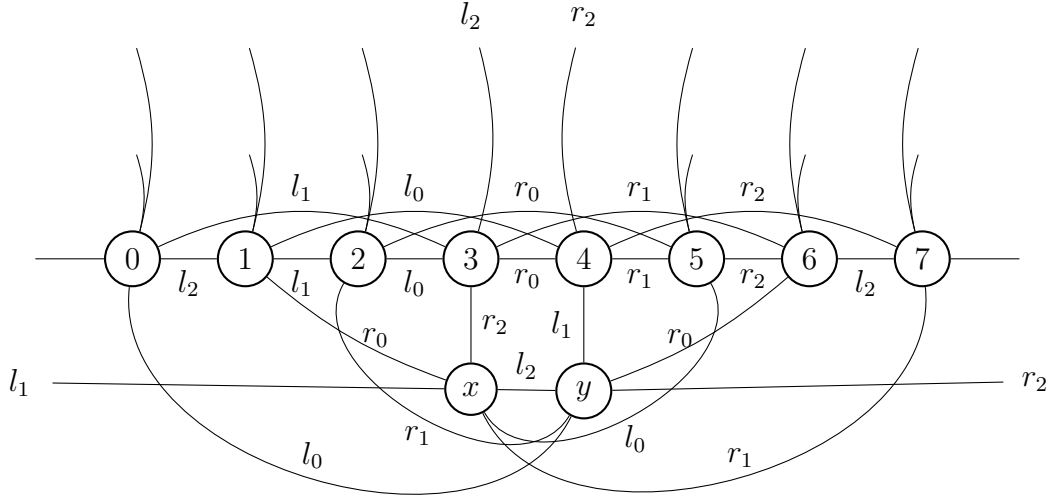
\begin{figure}[h]
    \centering
    \begin{tikzpicture}
    
    \foreach \x in {0,...,7}
    \node[vertex] (\x) at (1.5*\x,0) {\x};
    \foreach \x in {0,...,7}
    \node[vertex, draw=none] (t\x) [above=of \x] {};
    \foreach \x in {0,1,2,5,6,7}
    \node[vertex, draw=none] (tt\x) [above=of t\x] {};
     \node[vertex, draw=none] (tt3) [above=of t3] {$l_2$};
      \node[vertex, draw=none] (tt4) [above=of t4] {$r_2$};
    \node[vertex, draw = none] (-1) at (-1.5,0) {};
    \node[vertex, draw = none] (8) at (12,0) {};
   \node[vertex] (x) [below=of 3] {$x$};
   \node[vertex] (y) [below=of 4] {$y$};
   \node[vertex,draw=none] (l) [below=of -1] {$l_1$};
   \node[vertex,draw=none] (r) [below=of 8] {$r_2$};
    
    \path[-](3)  edge [align=left, right] node  {$r_2$} (x);
    \path[-](x)  edge [align=left, above] node  {$l_2$} (y);
    \path[-](y)  edge [align=left, left] node  {$l_1$} (4);
    \path[-](1)  edge [bend right=10,align=left, right] node  {$r_0$} (x);
    \path[-](6)  edge [bend left=10,align=left, left] node  {$r_0$} (y);
    \path[-](5)  edge [bend left=90,align=left, below] node  {$l_0$} (x);
    \path[-](2)  edge [bend right=90,align=left, below] node  {$r_1$} (y);
    \path[-](7)  edge [bend left=80,align=left, above] node  {$r_1$} (x);
    \path[-](0)  edge [bend right=80,align=left, above] node  {$l_0$} (y);
    \path[-](0)  edge [align=left, below] node  {$l_2$} (1);
    \path[-](1)  edge [ align=left, below] node  {$l_1$} (2);
    \path[-](2)  edge [ align=left, below] node  {$l_0$} (3);
    \path[-](3)  edge [ align=left, below] node  {$r_0$} (4);
    \path[-](4)  edge [ align=left, below] node  {$r_1$} (5);
    \path[-](5)  edge [ align=left, below] node  {$r_2$} (6);
    \path[-](6)  edge [ align=left, below] node  {$l_2$} (7);
    \path[-](7)  edge [ align=left, below] node  {} (8);
    \path[-](-1)  edge [ align=left, above] node  {} (0);
    \path[-](0)  edge [bend left=25, align=left, above] node  {$l_1$} (3);
    \path[-](1)  edge [bend left=25, align=left, above] node  {$l_0$} (4);
    \path[-](2)  edge [bend left=25, align=left, above] node  {$r_0$} (5);
    \path[-](3)  edge [bend left=25, align=left, above] node  {$r_1$} (6);
    \path[-](4)  edge [bend left=25, align=left, above] node  {$r_2$} (7);
    \foreach \x in {0,1,2}
    \path[-](\x)  edge [bend right=15, align=left, above] node  {} (t\x);
    \foreach \x in {7,6,5}
    \path[-](\x)  edge [bend left=15, align=left, above] node  {} (t\x);
    \foreach \x in {0,1,2,3}
    \path[-](\x)  edge [bend right=15, align=left, above] node  {} (tt\x);
    \foreach \x in {7,6,5,4}
    \path[-](\x)  edge [bend left=15, align=left, above] node  {} (tt\x);
    \path[-](l)  edge [align=left, right] node  {} (x);
    \path[-](r)  edge [align=left, right] node  {} (y);
    \end{tikzpicture}
    \caption{Extension of $\varphi$ on a part of $C_{12}(U_2)$. The color of each edge is written near the middle of the edge.}
    \label{fig:fstext}
\end{figure}

    \item Notice that each vertex of $C_{4m+4}(U_m)$ loses and recovers exactly one edge, except for $3m+3$ and $3m+4$ which remain unchanged. Assign the color of the lost edge to the new one. The two ends of a removed edge are at an odd distance on $C_{4m+4}(U_m)$, which ensures that exactly one is at an even distance of $m+1$ and exactly one is at an even distance of $m+2$. With the construction in step (\ref{step}), both ends get a new edge with a distinct vertex of ${x,y}$, so the coloring remains proper. Finally, assign $c(x,y)=l_m$, the only color $x$ and $y$ do not have in their edge neighbourhood (in Step (\ref{st:1}), no edge colored with $l_m$ is removed).
\end{enumerate}

This extension does not need to be added at this exact spot on the graph: before adding it, all vertices and colors have the same role in the construction. An extension only changes the color of edges of $4m$ vertices on the original graph, so on $C_{2(m+1)k}(U_m)$, multiple extensions can be added as long as there are always $4m$ vertices between two of them so they do not interfere with each other.

Now let $n=q(m+1)+r$ with $r\leq \min\left(m,\frac{q(m+1)}{2m}\right)$. Such integers $q$ and $r$ exist when $n\geq m(2m+1)$ An edge coloring of $C_{2n}(U_m)$ can be constructed by adding $r$ such $2$-vertices extensions to $C_{2q(m+1)}(U_m)$. \qed
\end{pf}

The colorings obtained in this way can be used to color edges of even length on $C_n([\![1,R]\!])$. For $k$, $p$, and $m$ integers $C_{2^pk}(\{2^p,3*2^p,...,(2m+1)2^p\})$ can be seen as the union of $2^p$ graphs isomorphic to $C_{2^pk}(U_m)$. Combining them with $\phi$ to color edges of odd length leads to the following result.

\begin{lem}\label{lem:las}
Let $R,k\in\mathbb{N}^*$.
\begin{itemize}
    \item If R is even, $C_{k(R+1)2^{1+\lceil\log R\rceil}}([\![1,R]\!])$ admits a $(2R+1)$-adjacent vertex-distinguishing edge-coloring;
    \item If R is odd, $C_{k(R+2)2^{1+\lceil\log R\rceil}}([\![1,R]\!])$ admits a $(2R+1)$-adjacent vertex-distinguishing edge-coloring.
\end{itemize}
\end{lem}

For an integer $d$ we write its $2$-adic valuation $\val(d)=\max\{p\in \Nb|2^p\text{ divides }d\}$. In the following we consider the partition of $[\![1,R]\!]$ according to the valuation of its elements, $[\![1,R]\!] = \bigcup\limits^{\lfloor\log R\rfloor}_{p=0}Q^p(R)$ where $Q^p(R)=\{d\in[\![1,R]\!]|\val(d)=p\}$. When the context is clear, we omit $R$ and only write $Q^p$.

\begin{pf}
Let $R\in\Nb$ and $q=\left\{\begin{array}{cc}
    (R+1) & \text{if $R$ is even} \\
    (R+2) & \text{if $R$ is odd}
\end{array}\right.$\\ We first consider the case $k=1$. We divide the edges of $C_{q2^{1+\lceil\log R\rceil}}([\![1,R]\!])$ according to their length with the partition of $[\![1,R]\!]$ given above and consider the graphs $C_{q2^{1+\lceil\log R\rceil}}(Q^p)$ for $p\in[\![0,\lfloor\log R\rfloor]\!]$.

$p=0$: If $R$ is even, $C_{q2^{1+\lceil\log R\rceil}}(Q^0)=C_{(2(\frac{R}{2}-1+1)+1)2^{1+\lceil\log R\rceil}}(U_{\frac{R}{2}-1})$, for which $\phi$ is a proper coloring in $R+1=2|Q^0|+1$ colors. If $R$ is odd, $C_{q2^{1+\lceil\log R\rceil}}(Q^0)=C_{(2(\frac{R-1}{2}+1)+1)2^{1+\lceil\log R\rceil}}(U_{\frac{R-1}{2}})$ for which $\phi$ is a proper edge-coloring in $R+2=2|Q^0|+1$ colors.

$p>0$: As $q\geq\frac R{2^p}$, the graph $C_{q2^{1+\lceil\log R\rceil}}(Q^p)$ has $2^p$ connected components, each of them isomorphic to $C_{q2^{1-p+\lceil\log R\rceil}}(U_{|Q^p|-1})$. If $|Q^p|=1$, then $\varphi$ is a proper coloring for each component. Otherwise, $q2^{\lceil\log R\rceil-p}\geq|Q^p|(2|Q^p|+1)$ and we can use Lemma \ref{lem:bet} to build a proper coloring for each component. By using the same $2|Q^p|$ colors for each connected component, we build a $2|Q^p|$-proper coloring for $C_{q2^{1+\lceil\log R\rceil}}(Q^p)$.

Color the edges of $C_{q2^{1+\lceil\log R\rceil}}([\![1,R]\!])$ with distinct colors for each set of edges so that the coloring remains proper. In the following, we use the super-index $^p$ for colors used on edges of length $d\in Q^p$. Lemma \ref{lem:str} ensures that adjacent vertices are distinguished by the coloring of edges of odd length (coloring of $C_{q2^{1+\lceil\log R\rceil}}(Q^0)$). The number of colors used is:
\begin{align}
    2|Q^0|+1+\sum\limits_{p=1}^{\lfloor\log R \rfloor}2|Q^p| &= 1+\sum\limits_{p=0}^{\lfloor\log R \rfloor}2|Q^p| \\
    &= 1+2\left|\bigcup\limits^{\lfloor\log R\rfloor}_{p=0}Q^p\right|\\
    &= 1+2R
\end{align}

For $k\geq2$, remind all colorings used to color $C_{q2^{1+\lceil\log R\rceil}}([\![1,R]\!])$ are periodic and can be applied on $C_{kq2^{1+\lceil\log R\rceil}}([\![1,R]\!])$.\qed
\end{pf}

Notice that we can specify exactly how each set of edges is colored. For edges of odd length, set the vertex $v^0$ which indices the coloring $\phi_{v^0}$. When $|Q^p|=1$, set $2^p$ vertices $(v_i^p)_i$ in distinct connected components of $C_{q2^{1+\lceil\log R\rceil}}(Q^p)$ which index the colorings $(\varphi_{v_i^p})_i$ used to color the components. When $|Q^p|>1$, specifying $2^p$ vertices $(v_i^p)_i$ does not set where the extensions are located. However, we can guarantee that the coloring on which they are added is $\varphi_{v_i^p}$ and that they are at distance at least $2|Q^p|$ of each $v_i^p$ on the connected component, so at distance at least $2^{p+1}|Q^p|$ of each $v_i^p$ on $C_{q2^{1+\lceil\log R\rceil}}(Q^p)$. For a sequence $W=\left(\left\{v_i^p|i\in[\![1,2^p]\!]\right\}\right)_{0\leq p\leq \log R}$ of sets of vertices such that for all $p$ and $i\neq j, v_i^p\neq v^p_j \mod 2^p$, we denote $\Phi_W$ a coloring obtained in this way. 

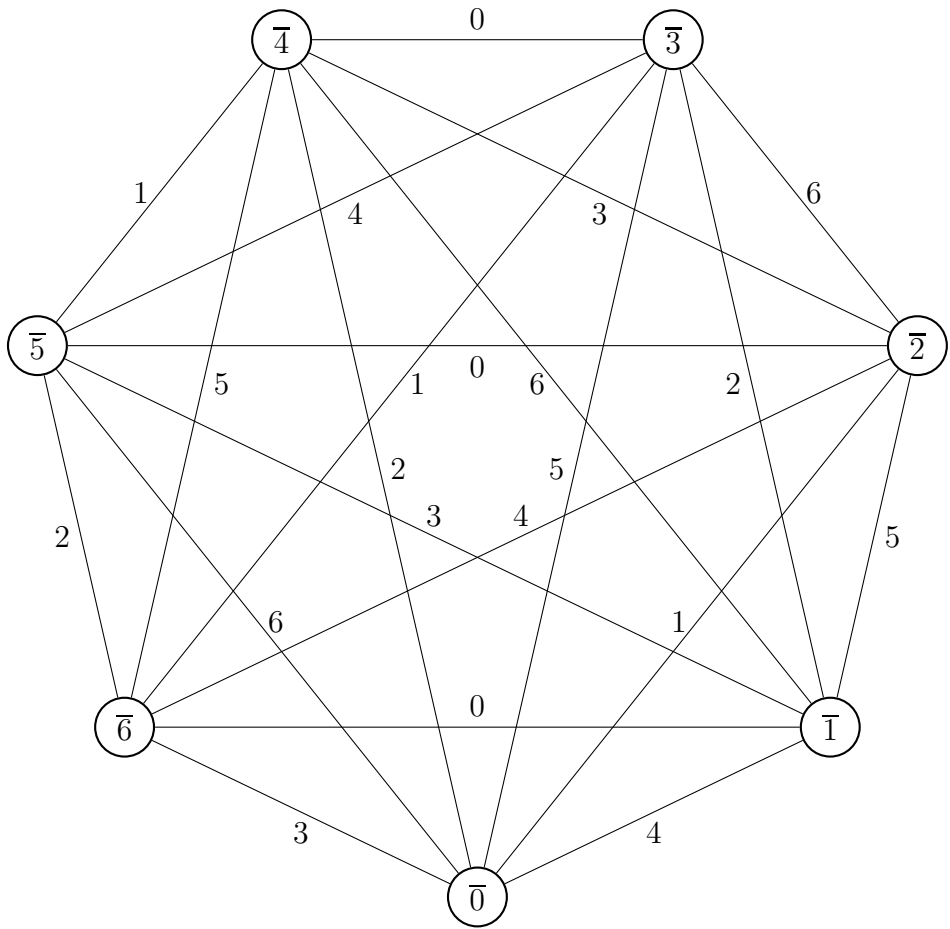
\begin{figure}[h]
    \centering
    \begin{tikzpicture}[scale=1]
  \foreach \x in {0,...,6}
    \node[vertex] (\x) at (\x*360/7-90:6) {$\overline\x$};
\foreach \x/\y in {6/1,4/3}
\path[-](\x)  edge [align=left, above] node  {$0$} (\y);
\path[-](5)  edge [align=left, below] node  {$0$} (2);
\foreach \x/\y in {6/0,4/2}
\path[-](\x)  edge [align=left, below] node  {$3$} (\y);
\path[-](5)  edge [align=left, above] node  {$3$} (1);
\foreach \x/\y in {1/0,5/3}
\path[-](\x)  edge [align=left, below] node  {$4$} (\y);
\path[-](6)  edge [align=left, above] node  {$4$} (2);
\foreach \x/\y in {3/2,5/0}
\path[-](\x)  edge [align=left, right] node  {$6$} (\y);
\path[-](4)  edge [align=left, left] node  {$6$} (1);
\foreach \x/\y in {6/5,3/1}
\path[-](\x)  edge [align=left, left] node  {$2$} (\y);
\path[-](4)  edge [align=left, right] node  {$2$} (0);
\foreach \x/\y in {0/2,4/5}
\path[-](\x)  edge [align=left, left] node  {$1$} (\y);
\path[-](6)  edge [align=left, right] node  {$1$} (3);
\foreach \x/\y in {6/4,1/2}
\path[-](\x)  edge [align=left, right] node  {$5$} (\y);
\path[-](3)  edge [align=left, left] node  {$5$} (0);

\end{tikzpicture}

    \caption{$\Psi_C$ on $K_{2R+1}$ for $R=3$ and $C=(0,1,2,3,4,5,6)$. The color of each edge is written near the middle of the edge.}
    \label{fig:complete}
\end{figure}

The next category of graphs, for which we want to find AVD colorings, are complete graphs of odd order: notice that for $R\in\Nb^*$, $K_{2R+1}=C_{2R+1}([\![1,R]\!])$. For a sequence of distinct colors $C=(c_i)_{1\leq i\leq 2R+1}$, let $\Psi_C$ be the coloring of $K_{2R+1}$ defined in the following way and illustrated in Figure \ref{fig:complete} for $R=3$:
\begin{itemize}
    \item Arrange the vertices of the graph $K_{2R+2}$ in the form of a regular $(2R+1)$-gon with one vertex in the center;
    \item Color the radial edges following the sequence $C$;
    \item Color each remaining edge with the color of the radial edge to which it is perpendicular;
    \item Remove the central vertex and the radial edges to get a colored $K_{2R+1}$ graph.
\end{itemize}
    
The proof that this coloring is vertex-distinguishing is made in \cite{BurrisA.C.1997VPE}. A direct consequence is the following property:

\begin{pr}\label{pr:cliq}
For any sequence of distinct colors $C=(c_i)_{1\leq i\leq 2R+1}$, the coloring $\Psi_C$ of $K_{2R+1}$ is a $(2R+1)$-adjacent vertex-distinguishing coloring. \qed
\end{pr}

\section{Fundamental Results}\label{sec:fun}

In Lemma \ref{lem:las}, we build an optimal adjacent vertex-distinguishing edge-coloring on $C_{mp}([\![1,R]\!])$ where $m$ is the period of the coloring and $p\in\Nb$. In order to prove Theorem \ref{thm:all}, we need to extend this construction to any large enough orders of circulant graphs. We do this by adding extensions to a colored $C_{mp}([\![1,R]\!])$, namely making the union of the colored graphs $C_{mp}([\![1,R]\!])$ and $C_{k}([\![1,R]\!])$, cutting each graph's edges between two vertices, and merging the two graphs by reuniting half-edges of the same color, resulting in a colored $C_{mp+k}([\![1,R]\!])$ graph. After showing how to add one extension, we compute for any $n$ the value of $p$ and how many extensions need to be added to $C_{mp}([\![1,R]\!])$ to get a colored $C_{n}([\![1,R]\!])$. We divide the proof of Theorem \ref{thm:all} according to the parity of $R$ and start with the following theorem when $R$ is even.

\begin{thm}\label{thm:even}
For $R$ even and $n\geq(R+1)2^{1+\lfloor \log R \rfloor}\left(R+(R+1)2^{\lfloor \log R \rfloor}\right)-2R$, $C_{n}([\![1,R]\!])$ admits a $(2R+1)$-adjacent vertex-distinguishing edge-coloring.
\end{thm}

\begin{pf}

Let $R$ be an even integer and $n\geq(R+1)2^{1+\lfloor \log R \rfloor}\left(R+(R+1)2^{\lfloor \log R \rfloor}\right)-2R$. Firstly, we show how to add one extension of $(2R+1)$ vertices to $C_{(R+1)2^{1+\lceil\log R\rceil}}([\![1,R]\!])$. Secondly, we compute for $k\in\Nb^*$ how many extensions can be added to $C_{k(R+1)2^{1+\lceil\log R\rceil}}([\![1,R]\!])$ and how to chose $k$ and $u$ so that $C_{n}([\![1,R]\!])$ is the result of adding $u$ extensions to $C_{k(R+1)2^{1+\lceil\log R\rceil}}([\![1,R]\!])$.

The two colored graphs which we consider are $C_{(R+1)2^{1+\lceil\log R\rceil}}([\![1,R]\!])$  and $K_{2R+1}$. The complete graph $K_{2R+1}$ is colored with $\Psi_C$, where:
 \begin{align}
        C=&(0^0,\label{6}\\
        &r^0_0, r^0_1, ... , r^0_{\frac{R}{2}-1},\label{7}\\
        &l^1_{|Q^1|-1}, l^2_{|Q^2|-1}, l^1_{|Q^1|-2}, l^3_{|Q^3|-1}, ..., l^{\val(R-2)}_0, l^{\val(R)}_0,\label{sequence}\\
        &r^{\val(R)}_0, ..., r^2_{|Q^2|-1}, r^1_{|Q^1|-1},\label{xseq}\\ 
        &l^0_{\frac{R}{2}-1}, ... , l^0_1, l^0_0)\label{10}
    \end{align}
    
    For $i\in[\![1,\frac R2]\!]$ the $i^\text{th}$ value in line (\ref{sequence}) is $l^{\val(2i)}_{|Q^{\val (2i)}|-\left\lceil\frac{2i}{2^{1+\val(2i)}}\right\rceil}$. Line (\ref{xseq}) mirrors line (\ref{sequence}) with all $l$ replaced by $r$.

To color the second graph, let $W^0=\{0\}$ and for $p\in[\![1,\lfloor \log R\rfloor]\!]$, let $W^p=[\![1+2^{p-1},3*2^{p-1}]\!]$. Consider the sequence $W=(W^p)_{0\leq p\leq \log R}$ and color $C_{(R+1)2^{1+\lceil\log R\rceil}}([\![1,R]\!])$ with a coloring $\Phi_W$.

We now consider the union of both graphs and sequence the vertices such that the sequence inside each component remains the same, and the vertices of the extension $K_{2R+1}$ are inserted between vertices $0$ and $1$ of $C_{(R+1)2^{1+\lceil\log R\rceil}}([\![1,R]\!])$, starting and ending with both ends of the edge of length $1$ colored with $0^0$. The sequence of vertices described by the color missing in their neighbourhood is the following (for the extension, start at the beginning of line \ref{xseq}, go through lines \ref{10}, \ref{6} and \ref{7} and finish at the end of line \ref{sequence} in the definition of $C$):
    \begin{align}
        ...,&\overline{0^0}, \overline{r^0_0}, ..., \mathbf{\overline{r^0_{\frac{R}{2}-1}}}, && \mathbf{\overline{r^{\val(R)}_0}}, ..., \overline{r^1_{|Q^1|-1}}, ..., \mathbf{\overline{l^{\val(R)}_0}}, && \mathbf{\overline{l^0_{\frac{R}{2}-1}}}, ... , \overline{l^0_0}, \overline{0^0},...\label{eq:sequence}\\
        &\text{Base graph} && \qquad\text{Extension} && \text{Base graph}\nonumber
    \end{align}

    We call \textit{junction vertices} the vertices which in this sequence follow or are followed by a vertex of the other component. They are in bold in Equation \ref{eq:sequence}. 

By changing the sequence of vertices we also change the length of some edges. In order to recover a circulant graph we cut each edge with length at least $R+1$ in two half-edges each incident to one end of the original edge. Cutting an edge $(a,b)$ results in two half edges, $(\underline a,b)$ incident to $a$ and $(a,\underline b)$ incident to $b$. We still write the vertices they are not incident to so that we can keep track of the edge they result from, notably to associate both half-edges and to remember their former color and length. We then show that for each vertex $v$, each half-edge incident to $v$ can be merged with a half-edge of the same color so that the new edges connect $v$ to all remaining vertices at distance at most $R$. The merging operation is shown in Figure \ref{fig:extension} for $R=2$.

    \begin{figure}[!h]
    \centering
    \begin{tikzpicture}[scale=0.8]
    
    \node[vertex] (0) at (0*360/25-127:10) {$\overline {l_0^0}$};
    \node[vertex] (1) at (1*360/25-127:10) {$\overline{0^0}$};
    \node[vertex] (2) at (2*360/25-127:10) {$\overline{r_0^0}$};
    \node[vertex] (3) at (3*360/25-127:10) {$\overline {l_0^0}$};
    \node[vertex] (4) at (4*360/25-127:10) {$\overline {0^0}$};
    \node[vertex] (5) at (5*360/25-127:10) {$\overline {r_0^0}$};
    
    \foreach \x in {0,5}
    \node[vertex, draw=none] (t\x) [above=of \x] {};
    \node[vertex, draw=none] (t1) [above=of 1] {$r^1_0$};
    \node[vertex, draw=none] (t4) [above=of 4] {$l^1_0$};
    \node[vertex, draw = none] (-1) at (-0.5*360/25-127:10) {};
    \node[vertex, draw = none] (6) at (5.5*360/25-127:10) {};
   
    \path[-](0)  edge [bend right=5, align=left, below] node  {$r_0^0$} (1);
    \path[-](1)  edge [bend right=5, align=left, below] node  {$l_0^0$} (2);
    \path[-](2)  edge [bend right=5, align=left, below left] node  {$0^0$} (3);
    \path[-](3)  edge [bend right=5, align=left, below] node  {$r_0^0$} (4);
    \path[-](4)  edge [bend right=5, align=left, below] node  {$l_0^0$} (5);
    
    \path[-](5)  edge [bend right=5, align=left, below] node  {} (6);
    \path[-](-1)  edge [bend right=5, align=left, above] node  {} (0);
    
    \path[-](0)  edge [bend left=15, align=left, above right] node  {$l_0^1$} (2);
    \path[-](1)  edge [bend left=15, align=left, above] node  {$l_0^1$} (3);
    \path[-](2)  edge [bend left=15, align=left, above] node  {$r_0^1$} (4);
        \path[-](3)  edge [bend left=15, align=left, above left] node  {$r_0^1$} (5);
    \foreach \x in {0,1}
    \path[-](\x)  edge [bend right=15, align=left, above] node  {} (t\x);
    \foreach \x in {4,5}
    \path[-](\x)  edge [bend left=15, align=left, above] node  {} (t\x);

    \node[vertex] (b0) at ($(-90:2.3)+(0,-14)$) {$\overline{0^0}$};
    \node[vertex] (b1) at ($(-18:2.3)+(0,-14)$) {$\overline{r_0^0}$};
    \node[vertex] (b2) at ($(54:2.3)+(0,-14)$) {$\overline{l_0^1}$};
    \node[vertex] (b3) at ($(126:2.3)+(0,-14)$) {$\overline{r_0^1}$};
    \node[vertex] (b4) at ($(-162:2.3)+(0,-14)$) {$\overline{l_0^0}$};
    
    \foreach \x/\y in {b4/b1}
\path[-](\x)  edge [align=left, below] node  {$0^0$} (\y);
\path[-](b2)  edge [align=left, above right] node  {$0^0$} (b3);
\foreach \x/\y in {b4/b0}
\path[-](\x)  edge [align=left, below] node  {${l_0^1}$} (\y);
\path[-](b3)  edge [align=left, above right] node  {${l_0^1}$} (b1);
\foreach \x/\y in {b3/b0}
\path[-](\x)  edge [align=left, left] node  {${l_0^0}$} (\y);
\path[-](b2)  edge [align=left, right] node  {${l_0^0}$} (b1);
\foreach \x/\y in {b2/b0}
\path[-](\x)  edge [align=left, right] node  {${r_0^0}$} (\y);
\path[-](b4)  edge [align=left, left] node  {${r_0^0}$} (b3);
\foreach \x/\y in {b0/b1}
\path[-](\x)  edge [align=left, below] node  {${r_0^1}$} (\y);
\path[-](b2)  edge [align=left, above left] node  {${r_0^1}$} (b4);

\foreach \x/\y in {0/$\overline {l_0^0}$,1/$\overline{0^0}$,2/$\overline{r_0^0}$,3/$\overline {l_0^0}$,4/$\overline {0^0}$,5/$\overline {r_0^0}$}
\node[vertex] (2\x) at ($(\x*360/25-127:10)+(0,-13)$) {\y};
\foreach \x/\y in {0/$\overline {0^0}$,1/$\overline{r_0^0}$,2/$\overline{l_0^1}$,3/$\overline {r_0^1}$,4/$\overline {l^0_0}$}
\node[vertex] (c\x) at ($(\x*72-90:2.3)+(0,-28)$) {\y};
\foreach \x in {0,5}
    \node[vertex, draw=none] (t2\x) [above=of 2\x] {};
    \node[vertex, draw=none] (t21) [above=of 21] {$r^1_0$};
    \node[vertex, draw=none] (t24) [above=of 24] {$l^1_0$};
    \node[vertex, draw = none] (-21) at ($(-0.5*360/25-127:10)+(0,-13)$) {};
    \node[vertex, draw = none] (26) at ($(5.5*360/25-127:10)+(0,-13)$) {};

\foreach \x/\y in {c3/c0,c2/c1}
\path[-](\x)  edge [align=left, right] node  {${l_0^0}$} (\y);
\foreach \x/\y in {c3/c4,c2/c0}
\path[-](\x)  edge [align=left, left] node  {${r_0^0}$} (\y);

 \path[-](20)  edge [bend right=5, align=left, below] node  {$r_0^0$} (21);
    \path[-](21)  edge [bend right=5, align=left, below] node  {$l_0^0$} (22);
    \path[-](23)  edge [bend right=5, align=left, below] node  {$r_0^0$} (24);
    \path[-](24)  edge [bend right=5, align=left, below] node  {$l_0^0$} (25);
    \foreach \x in {20,21}
    \path[-](\x)  edge [bend right=15, align=left, above] node  {} (t\x);
    \foreach \x in {24,25}
    \path[-](\x)  edge [bend left=15, align=left, above] node  {} (t\x);
    \path[-](25)  edge [bend right=5, align=left, below] node  {} (26);
    \path[-](-21)  edge [bend right=5, align=left, above] node  {} (20);
    \path[-](20)  edge [bend left=20, align=left, above right] node  {$l_0^1$} (22);
    \path[-](23)  edge [bend left=20, align=left, above left] node  {$r_0^1$} (25);
    \path[-](c4) edge [below] node {$0^0$} (c1);
    \path[-](21)  edge [align=left, right] node  {${l_0^1}$} (c3);
    \path[-](24)  edge [align=left, left] node  {${r_0^1}$} (c2);
    \path[-](22)  edge [align=left, right] node  {${0^0}$} (c3);
    \path[-](23)  edge [align=left, left] node  {${0^0}$} (c2);
    \path[-](c0)  edge [align=left, below] node  {${l_0^1}$} (c4);
    \path[-](c0)  edge [align=left, below] node  {${r_0^1}$} (c1);
    \path[-](22)  edge [bend right=45, align=left, right] node  {${r_0^1}$} (c4);
    \path[-](23)  edge [bend left=45, align=left,  right] node  {${l_0^1}$} (c1);
    
    \draw [dash pattern=on 3pt off 3pt] (0,-8) -- (0,-14);
    \end{tikzpicture}
    \caption{The extension before and after the cut for $R=2$. The color of each edge is written near the middle of the edge.}
    \label{fig:extension}
\end{figure}
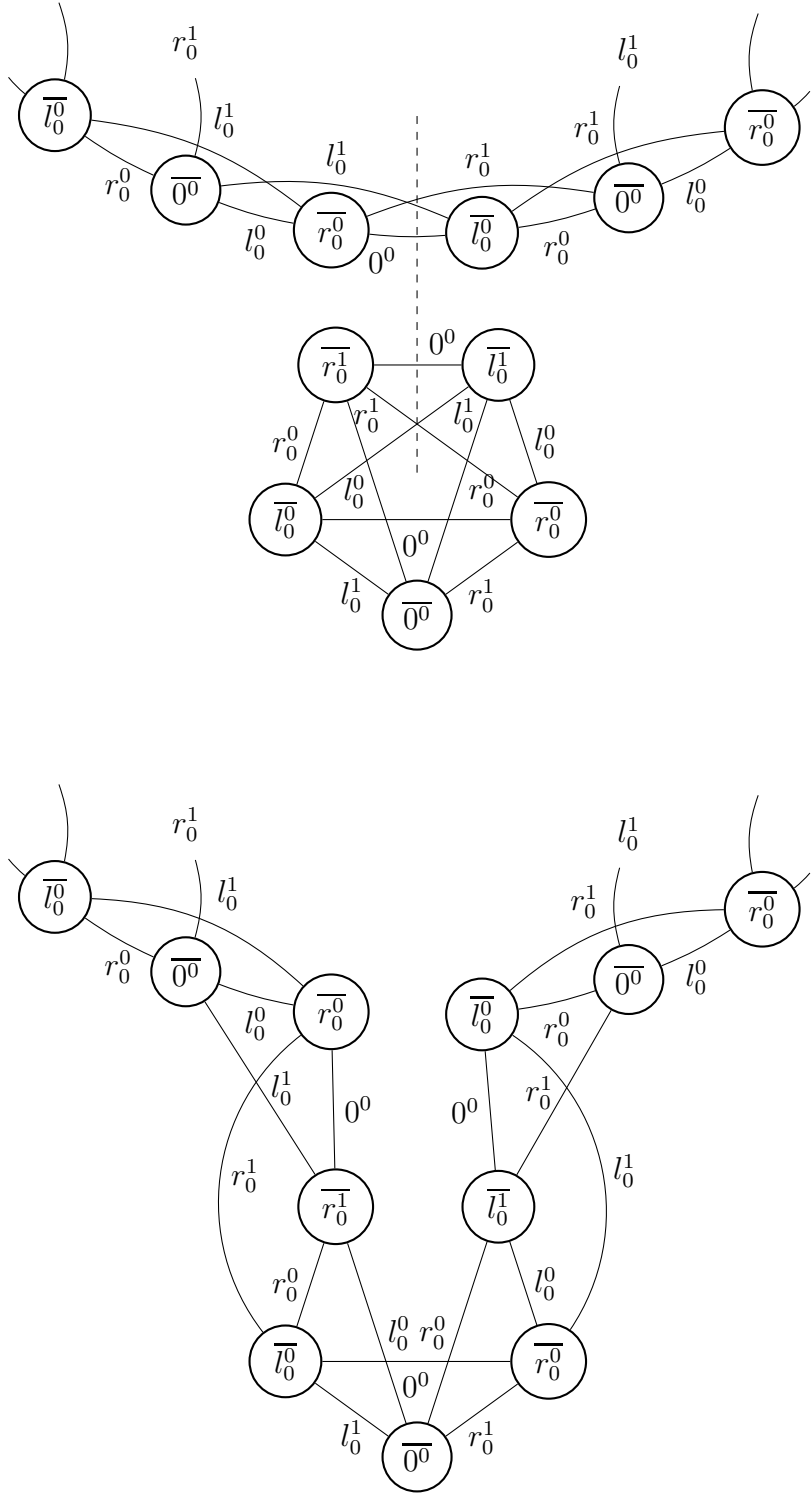

This will prove that the extension operation does not change the set of colors and lengths of incident edges of any vertex, so the result is a proper edge coloring of $C_{2R+1+(R+1)2^{1+\lceil\log R\rceil}}([\![1,R]\!])$. Its adjacent vertex-distinguishing property can then be seen by observing in Equation (\ref{eq:sequence}) that we only add to the sequence of incident edge color sets one period and $R$ color sets distinct with one another and with the other color sets.

For each vertex $v$, let $H_v=\{(a,\underline v) | v\text{ is stricly closer than } a \text{ to a junction vertex}\}$. We denote by $H_v^c$ the set of colors of the half-edges in $H_v$. We are going to show that for two vertices $u$ on the extension and $v$ on the base graph at the same position relatively to the cuts, $H_u^c=H_v^c$. The same is true for the set of lengths of the cut edges $H^l_u$ and $H^l_v$. Thus we can merge the half-edges of $H_u$ to the complementary half-edges to the ones of $H_v$. The same way, by merging half-edges resulting from edges colored with $0^0$ for which both ends are at the same distance from the junction vertices, each vertex recovers edges of each length and each color that it lost from cutting edges.

We consider the vertices before the cut on the base graph (on the left in Figure \ref{fig:extension}). The first vertex $v$ for which $H_v\neq\emptyset$ is $1-\frac R2$ (the vertex at distance $\frac R2-1$ of the junction vertex $0$). Indeed, a half-edge is in $H_{-\frac R2}$ if the distance between $1$ and the other end $b$ of the corresponding edge is greater than the distance between $-\frac R2$ and $0$. This means $b>\frac R2+1$, so the distance between $-\frac R2$ and $b$ is greater than $R+2$, and the edge $(-\frac R2,b)$ does not exist in $C_{(R+1)2^{1+\lceil\log R\rceil}}([\![1,R]\!])$. The half-edge in $H_{1-\frac{R}{2}}$ is obtained by cutting its edge of length $R$ colored with $r^{\val(R)}_0$:

\begin{align}
    \Phi_W(1-\frac{R}{2},1+\frac{R}{2})&=(1-2^{\val(R)-1},1+2^{\val(R)-1})\\
    &=r^{\val(R)}_0 \\
    & \text{as}\quad \Phi_W(1-2^{\val(R)-1}-2^{\val(R)},1-2^{\val(R)-1})=l^{\val(R)}_0
\end{align}

We now show that if a vertex $u$ is between a vertex $v$ and the cut then $H_u^c\subseteq H_v^c$. Let $d\in[\![1,R]\!]$ and $c$ the color of $(\underline{-j}, d-j)\in H_{-j}$. If $d\in Q^p$, then in the base graph, $c$ is the color of $2^p$ consecutive edges of length $d$. Let $o$ be the rank (starting from $0$) of vertex $-j$ in this sequence. Let $i\leq j$ and let $q$ and $r<2^p$ be the integers such that $o+i=2^pq+r$. The integer $r$ is the rank of vertex $i-j$ in the sequence of $2^p$ rightbound edges of color $c$ and length $d-q2^{p+1}$. Thus the edge $(i-j,i-j+d-2^{p+1}q)$ is colored with $c$. We now show $(\underline{i-j},i-j+d-2^{p+1}q)\in H_{i-j}$ by proving that $j-i<i-j+d-2^{p+1}q-1 \quad(*)$.

If $p=0$ then $q=i$ and $2^{p+1}=2$. Thus we have $(\underline{-j}, d-j)\in H_{-j}\implies 2j+1<d$, which proves $(*)$. Otherwise, let $q_j$ and $q_d$ be the integers such that $j+o=2^{p-1}(2q_j+1)-1$ and $d=2^p(2q_d+1)$. We have $o<2^p$ so $j>2^{p-1}(2q_j-1)$, which allows us to write the following implications between $d>2j+1$ and $(*)$:
\begin{align}
    d&>2j+1\\
    2^p(2q_d+1)&>2^p(2q_j-1)+1\nonumber\\
    q_d&>q_j-1+2^{-p-1}\nonumber\\
    q_d&\geq q_j\nonumber\\
    2^p(2q_d+1)-2^p(2q_j+1)+1&>0\nonumber\\
    d-2(j+o)-1&>0\nonumber\\
    d-j-2(2^pq+r-i)-1&> j\nonumber\\
    d-2^{p+1}q+i-j-1&> j-i \tag{*}
\end{align}

For $j\in[\![1,\frac R2 -1]\!]$, we have $|H_{1-j}|-|H_{-j}|=2$. The additional colors in $H^c_{1-j}$ are:
\begin{itemize}
    \item the color $r^0_{\frac R2-1-j}$ from the edge of length $R-1$:
    \begin{align}
        \Phi_W(1-j,R-j)=\Phi_W(\frac R2-j,\frac R2-j+1)=r^0_{\frac R2-1-j}
    \end{align}
 \item the $(\frac R2-j)^\text{th}$ color in the sequence of colors of line (\ref{xseq}). If $j+1=2^{p-1}(2q+1)$, this color is $r^p_{|Q^p|-q-1}$, and it comes from the edge of length $2^p(2|Q^p|-1)$):
\begin{align}
        &\Phi_W(2-2^{p-1}(2q+1),2-2^{p-1}(2q+1)+2^p(2|Q^p|-1)\\
        =&\Phi_W(2-2^{p-1}(2q+1)+2^p(|Q^p|-1),2-2^{p-1}(2q+1)+2^p|Q^p|)\nonumber\\
        =&\Phi_W(2^p(|Q^p|-q-1)+2-2^{p-1},2^p(|Q^p|-q-1)+2+2^{p-1})\nonumber\\
        =&r^p_{|Q^p|-q-1}
    \end{align}
\end{itemize}

Next we prove $H^c_{-j}=H^c_{v_j}$ where $v_j$ is the vertex of the extension at the $j^{\text{th}}$ position before the second junction vertex of the extension (on the right of the cut in Figure \ref{fig:extension}). Again, for $i\in[\![0,j]\!]$, $H^c_{v_j}\subseteq H^c_{v_{j-i}}$ as if $c\in H^c_{v_j}$ comes from an edge of former length $d$, then the edge of length $d-2i$ crossing the cut is also colored with $c$ and generates a half edge in $H^c_{v_{j-i}}$. The set $H_{v_{\frac{R}{2}-1}}$ is also the first non-empty set of half-edges, as it includes the half-edge obtained from an edge of length $R$ and colored with $r^{\val(R)}_0$, the missing color of the first junction vertex (the vertex at distance $\frac R2$ of both ends of the edge of length $R$). The colors in $H^c_{v_{j-1}}\setminus H^c_{v_j}$ are:
\begin{itemize}
    \item the missing color of the vertex at distance $\frac R2$ from $v_{j-1}$ towards the cut (the $(\frac R2-j)^\text{th}$ color in the sequence (\ref{xseq})), which is coloring the half-edge obtained from its edge of length $R$;
    \item the color of the edge between the vertices at distances $\frac R2-j-1$ and $\frac R2-j$ away from the left junction vertex ($r^0_{\frac R2-j-1})$, which is coloring the half-edge obtained from its edge of length  $R-1$. 
\end{itemize} 

We proved $H^c_{-j}=H^c_{v_j}$. We also have $H^l_{-j}=H^l_{v_j}=[\![2(j+1),R]\!]$. Notice that vertex $d$, which is at distance $j+d$ of vertex $-j$ on the base graph, is at distance $j+d$ of $v_j$ on the extended graph. Thus if $(\underline{-j},d)\in H_{-j}$, we can merge $(\underline{-j},d)$ with the half-edge of $H_{v_j}$ of the same color, and this way give an edge of each length in $H^l_{-j}$ to vertex $-j$. The same can be done to merge the half-edges of $H_{-j}$. Also, it can be shown the same way for the vertices on the other side of the junction vertices. When cutting edges, there are two possibilities: either each end is at the same distance from the junction vertices, or one is closer and the other one is further. Thus after cutting the edges there are exactly as many half-edges incident to a vertex closer to the junction vertices as there are incident to a vertex further from the junction vertices. Notice that we merge all the edges from the first category to distinct half-edges in the second category, while keeping the length of the half-edge in the second category.

The only remaining half-edges are obtained by cutting an edge with both ends at the same distance of the junction vertices. These are all colored with $0^0$ and can be merged with one another to recover their original length. We have shown that for each vertex, we could merge all half-edges incident to this vertex with half-edges of the same colors to recover edges of each missing length. The resulting colored graph is $C_{2R+1+(R+1)2^{1+\lceil\log R\rceil}}([\![1,R]\!])$ and the coloring is adjacent vertex-distinguishing.

In this construction all colors attributed to edges of length in one set $Q^p$ play the same role and can be interverted. In order to be able to place an extension, the cut on the base graph must be between two vertices $v$ and $v+1$, where for all $p\in[\![1,\lfloor\log R\rfloor]\!]$, the rightbound edges of length $2^p$ on the vertices between the vertex at distance $2^{p-1}-1$  on the left and at distance $2^{p-1}$ on the right of $v$ have the same color. This situation appears for edges of length $2^p$ every $2^p$ edges, and as it is the case for all $p$ on vertex $0$, it is also the case for all $p$ on vertices $2^{\lfloor\log R\rfloor}k$ for all $k$. Also, the extension modifies only the rightbound (resp. leftbound) edges of the $\frac R 2 \leq 2^{\lfloor\log R\rfloor}$ vertices on the left (resp. right) of the cut so an extension can be placed on any of these positions. Therefore, it is possible to add up to $k$ extensions to $C_{k2^{\lfloor\log R\rfloor}}([\![1,R]\!])$.

We have $$\gcd(2R+1,(R+1)2^{1+\lfloor\log R\rfloor})=1$$
Therefore, $$\forall i<(R+1)2^{1+\lfloor\log R\rfloor},\: \exists\:!u<(R+1)2^{1+\lfloor\log R\rfloor}$$
such that $$u(2R+1)=i\mod{(R+1)2^{1+\lfloor\log R\rfloor}}$$
So $$\forall z\in\mathbb Z, \exists!\: u<(R+1)2^{1+\lfloor\log R\rfloor} \text{ and } v\in\mathbb Z$$
such that $$z=u(2R+1)+v(R+1)2^{1+\lfloor\log R\rfloor}$$
If $u\leq2v(R+1)$ then a coloring of $C_z([\![1,R]\!])$ can be obtained by adding $u$ extensions of $2R+1$ vertices to $C_{v(R+1)2^{1+\lfloor\log R\rfloor}}([\![1,R]\!])$. The largest value of $z$ for which $u>2v(R+1)$ is : $$z_{max}=(R+1)2^{1+\lfloor \log R \rfloor}\left(R+(R+1)2^{\lfloor \log R \rfloor}\right)-(2R+1)$$ 
So as $n\geq z_{max}+1$,
there exists a $(2R+1)$-adjacent vertex-distinguishing edge-coloring of $C_n([\![1,R]\!])$. \qed
\end{pf}
\begin{thm}\label{thm:odd} Let $R,n\in\mathbb{N}^*$
\begin{itemize}
    \item If $n\geq(R+2)2^{1+\lfloor \log R \rfloor}\left(R+(R+2)2^{\lfloor \log R \rfloor}\right)-2R$ and $R=3 \mod{6}\text{ or } R=5\mod{6}$, $C_{n}([\![1,R]\!])$ admits a $(2R+1)$-adjacent vertex-distinguishing edge-coloring;
    \item If $R=1\mod{6}$ and $n\geq\frac{R+2}32^{1+\lfloor \log R \rfloor}\left(R+\frac{R+2}32^{\lfloor \log R \rfloor}\right)-2R$, $C_{3n}([\![1,R]\!])$ admits a $(2R+1)$-adjacent vertex-distinguishing edge-coloring.
\end{itemize}
\end{thm}

\begin{pf}

The proof of Theorem \ref{thm:odd} is the same as the proof of Theorem \ref{thm:even} except the extension is slightly different and if $R=1\mod{6}$, $2R+1$ is not prime with the period of the coloring on the base graph. In this case, the extension is not sufficient to construct $C_{n}([\![1,R]\!])$ for all values of $n$. The base graph is still colored with $\Phi_W$, where $W$ is defined in the same way as for $R$ even. The sequence $C$ which is used in $\Psi_C$ to color $K_{2R+1}$ is:

    \begin{align}
        C=&(0^0,\label{eq:6}\\
        &r^0_0, r^0_1, ... , r^0_{\frac{R-1}{2}},\label{eq:7}\\
        &l^1_{|Q^1|-1}, l^2_{|Q^2|-1}, l^1_{|Q^1|-2}, l^3_{|Q^3|-1}, ..., l^{\val(R-3)}_0, l^{\val(R-1)}_0,\label{eq:odsequence}\\
        &r^{\val(R-1)}_0, ..., r^2_{|Q^2|-1}, r^1_{|Q^1|-1},\label{eq:odxseq}\\ 
        &l^0_{\frac{R-1}{2}}, ... , l^0_1, l^0_0)\label{eq:10}
    \end{align}

The proof that $H^c_{-j}\subseteq H^c_{1-j}$ is still valid. Before the cut on the base graph, the last vertex $v$ for which $H_v=\emptyset$ is $\frac{1-R}{2}$. For $j\in [\![1,\frac{R-1}{2}]\!]$, the colors in $H^c_{-j} \setminus H^c_{1-j}$ are:
\begin{itemize}
    \item $r^0_{j-1}$ (edge of length $R$);
    \item and the $(\frac{R+1}{2}-j)^\text{th}$ color in the sequence \ref{eq:odsequence} (if this color is $l^p_k$, then it is lost by cutting an edge of length $2^p(1+2|Q^p|-1)$).
\end{itemize}  

Before the cut on the extension, $H_{v_{\frac{R-1}{2}}}=\emptyset$ and for $j\in [\![1,\frac{R-1}{2}]\!]$, the colors in $H^c_{v_{-j}} \setminus H^c_{v_{1-j}}$ are $r^0_{j-1}$ (edge of length $R$) and the $(\frac{R+1}{2}-j)^\text{th}$ color in the sequence \ref{eq:odsequence} with its edge of length $R-1$.

Again, it is possible to add up to $k$ extensions to $C_{k2^{\lfloor\log R\rfloor}}([\![1,R]\!])$.

We have $$\gcd(2R+1,(R+2)2^{1+\lfloor\log R\rfloor})=\gcd(R-1,3)$$
For $R\neq1\mod3$ and $n\geq(R+2)2^{1+\lfloor \log R \rfloor}\left(R+(R+2)2^{\lfloor \log R \rfloor}\right)-2R$, there exists $u$ and $v$ such that $$n=u(2R+1)+v(R+2)2^{1+\lfloor\log R\rfloor}$$ and $C_n([\![1,R]\!])$ can be obtained by adding $u$ $(2R+1)$-vertices extensions to $C_{v(R+2)2^{1+\lfloor\log R\rfloor}}([\![1,R]\!])$. 

For $R=1\mod3$ and $n\geq\frac{R+2}32^{1+\lfloor \log R \rfloor}\left(R+\frac{R+2}32^{\lfloor \log R \rfloor}\right)-2R$, there exists $u$ and $v$ such that $$3n=u(2R+1)+v(R+2)2^{1+\lfloor\log R\rfloor}$$ Add $u$ $(2R+1)$-extensions to $C_{v(R+2)2^{1+\lfloor\log R\rfloor}}([\![1,R]\!])$ to get a coloring for $C_{3n}([\![1,R]\!])$. \qed
\end{pf}

\section{Perspectives}\label{sec:per}

In this paper we proved Theorem \ref{thm:all} which states that $\chi^{\prime}_a(C_n([\![1,R]\!]))=2R+1$ for most values of $R$ and $n$. For values of $R$ and $n$ which are not covered by Theorem \ref{thm:all}, the value of $\chi^{\prime}_a(C_n([\![1,R]\!]))$ is unknown. Mainly, the following question remains.

\begin{qu}
Let $R=1\mod6$ and $n\neq0\mod3$. Does there exist a $(2R+1)$-adjacent vertex-distinguishing edge-coloring of $C_n([\![1,R]\!])$?
\end{qu}

    For $n\neq0\mod3$ it is well known that $\chi^{\prime}_{a}(C_n(\{1\}))=4$ as $C_n(\{1\})$ is a cycle which admits a $3$-adjacent vertex-distinguishing edge-coloring if and only if $n=0\mod3$. For $R\geq7$ it is unclear whether a $(2R+1)$-adjacent vertex-distinguishing edge-coloring exists for $n\neq0\mod3$. A way to prove its existence would be to find a extension of $k$-vertices where $k\neq0\mod3$ which could be added once or twice to any $C_{3m}([\![1,R]\!])$ colored in the way we defined to get a coloring of $C_{n}([\![1,R]\!])$. However, such an extension does not necessarily exist.
    
    Another question is for any $R$ and $n\leq(R+2)2^{1+\lfloor \log R \rfloor}\left(R+(R+2)2^{\lfloor \log R \rfloor}\right)-2R$ whether the theorem holds. The property is not true for all $n$ and $R$. For example, $\chi^{\prime}_{a}(C_7(\{1,2\})=6$, as shown in \cite{Mockov}. However, the authors prove that for any $n\geq5$ and $n\neq7$, $\chi^{\prime}_{a}(C_n(\{1,2\})=5$. As a comparison for $R=2$ Theorem \ref{thm:all} only covers $n\geq68$ (the coloring we defined can be still be applied for $29$ out of $63$ values of $n\in[\![5,67]\!]$).

An interesting idea in the proof of Theorem \ref{thm:all} is highlighted by Lemma \ref{lem:str} which introduces a new type of edge-coloring that can be defined in the following way.

\begin{defn}
Let $G=(V,E)$ and $G^\prime=(V,E^\prime)$ be two graphs on the same set of vertices, and $\varphi:E\rightarrow\varphi(E)$ be an edge-coloring of $G$. We say $\varphi$ is $(G,G^\prime)$-vertex-distinguishing if $\varphi$ is a proper edge-coloring of $G$ and for all $u,v\in V$ such that $(u,v)\in E^\prime$, $\varphi(N_G(u))\neq\varphi(N_G(v))$.
\end{defn}

Lemma \ref{lem:str} could be written as \textit{$\phi$ is a $(C_{(2m+3)k}(\{1,3...2m+1\}),C_{(2m+3)k}([\![1,2m+2]\!])$-vertex-distinguishing coloring}. This definition unifies the concepts of adjacent vertex-distinguishing edge-colorings when $G^\prime=G$, strong edge-colorings when $G^\prime=K_{|V|}$, and $d$-strong edge-colorings when $G^\prime$ is a $d$-power of $G$. Observe that in the proof of Lemma \ref{lem:las} we color the edges of $G=C_{(2m+3)k}(\{1, 3, ... ,\\2m+1\})$ with a $(G,G^\prime)$-vertex-distinguishing coloring where $G^\prime=C_{(2m+3)k}([\![1,\\2m+2]\!])$, and we also color the edges of $G^\prime \setminus G$ with a proper coloring using colors distinct from the ones used on $G$. This concept introduces a new problem for edge-colorings of graphs.

\begin{prob}
Given $k\in\Nb$, $G=(V,E)$ and $G^\prime=(V,E^\prime)$, does there exist a $(G,G^\prime)$-vertex-distinguishing coloring using at most $k$ colors?
\end{prob}

\bibliography{main}


\end{document}